\documentclass[prl,twocolumn,super,showpacs,groupedaddress]{revtex4}

\usepackage{graphicx}
\usepackage{amssymb}
\usepackage{amsmath}
\usepackage{bbm}
\begin{document}

\title{Emergent comet-like swarming of optically driven thermally active colloids}

\author{Jack A. Cohen}
 % \email[]{j.cohen@physics.ox.ac.uk}
%   \affiliation{Department of Physics, University of Warwick, Coventry, United Kingdom}
   \affiliation{Rudolf Peierls Centre for Theoretical Physics, University of Oxford, Oxford, United Kingdom}
\author{Ramin Golestanian}
%\email[]{ramin.golestanian@physics.ox.ac.uk}
   \affiliation{Rudolf Peierls Centre for Theoretical Physics, University of Oxford,
Oxford, United Kingdom}
\date{\today}

\begin{abstract}
We propose a simple system of optically driven colloids that convert light into heat and move in response to self- and collectively- generated thermal gradients. We show that the system exhibits self-organization into a moving comet-like swarm and characterize the structure and response of the swarm to a light intensity dependent external tuning parameter. We observe many interesting features in this nonequilibrium system including circulation and evaporation, intensity-dependent shape, density and temperature fluctuations, and ejection of hot colloids from the swarm tip.
\end{abstract}

\maketitle

Hierarchical self-organization of active components that convert energy into functional motile structures is one of the defining elements of living matter. This concept has motivated a decade of theoretical studies \cite{JoPr2009,Ramaswamy2010,MCMetal2013} that has culminated in a recent surge of well-controlled {\it in vitro} experiments \cite{Bausch2010,Dogic2012,Nagai2012}. The ubiquity of active self-organization in nature---from bird flocks \cite{Cavagna2010} to bacterial suspensions \cite{Ray}---and the rich fundamental properties exhibited by systems with such capability \cite{Vicsek1995,TonerTu1995,Chate2004} suggest that understanding such nonequilibrium phenomena could also help us develop a new paradigm in engineering by designing emergent behavior. The advent of active colloids \cite{paxton2004,Howse2007,Jiang2010} provides simple synthetic building blocks that could be used in bottom-up studies of the relationship between built-in functionalities of the modules and their resulting collective behaviors \cite{Theurkauff2012,NYU2013}.

Colloids can be driven into motion by gradients in chemical, electrostatic, or thermal fields that may exist externally to the colloid \cite{Anderson1989}. By tailoring the surface activity of the colloids, it is possible to produce such fields natively and generate self-propulsion \cite{gla2005,ruckner2007}. A collection of such interacting active colloids could serve as a promising model system to study collective non-equilibrium dynamics \cite{Vicsek2012}, as both the single-particle activity and the interactions could be controlled by construction. Here we consider a simple system of active colloids that receive energy by surface absorption of light. The colloids take advantage of the natural asymmetries in the system to create nonequilibrium conditions that drive them into a range of collective behavior, and in particular, self-organization into a moving comet-like swarm (see Fig. 1a) with novel nonequilibrium dynamics. We observe persistent circulation flow within the swarm (see Fig. 1b), evaporation, ejection of hot colloids from the head of the swarm and large shape fluctuations that induce fission. The rich behavior of the dynamic comet-like swarm can be controlled by a single external parameter proportional to the intensity of illumination.

\begin{figure}[htbp]
\centering
\includegraphics[width=\linewidth]{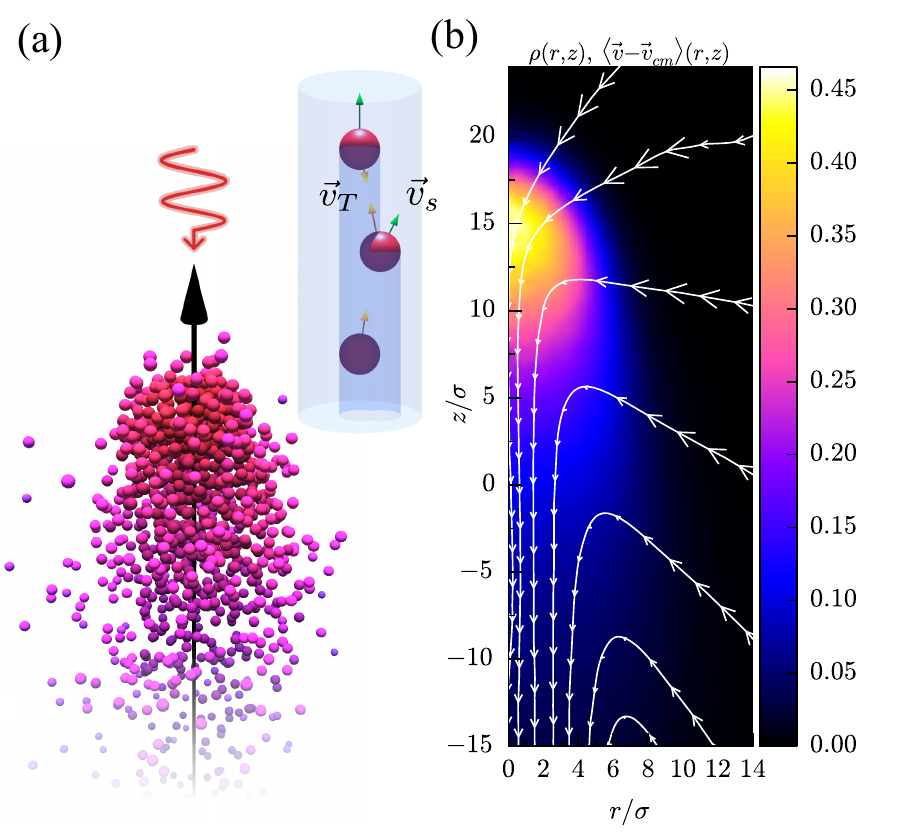}
\caption{
(a) Snapshot of the comet-like swarm with the light incident along $-\hat{z}$ with an inset schematic of the shadowing showing the collectively generated velocity $\vec{v}_T$ and the self-generated propulsion velocity $\vec{v}_s$. (b) Axially and time-averaged density field in the swarm center of mass frame presented in cylindrical coordinates $(r,z)$ overlaid with the relative average colloid velocity streamlines showing circulation, with arrow size representing the magnitude, generated from $\eta=10$ with $N_0=1024$.\label{fig:fig1}}
\end{figure}

\begin{figure*}[htbp]
\centering
\includegraphics[width=\linewidth]{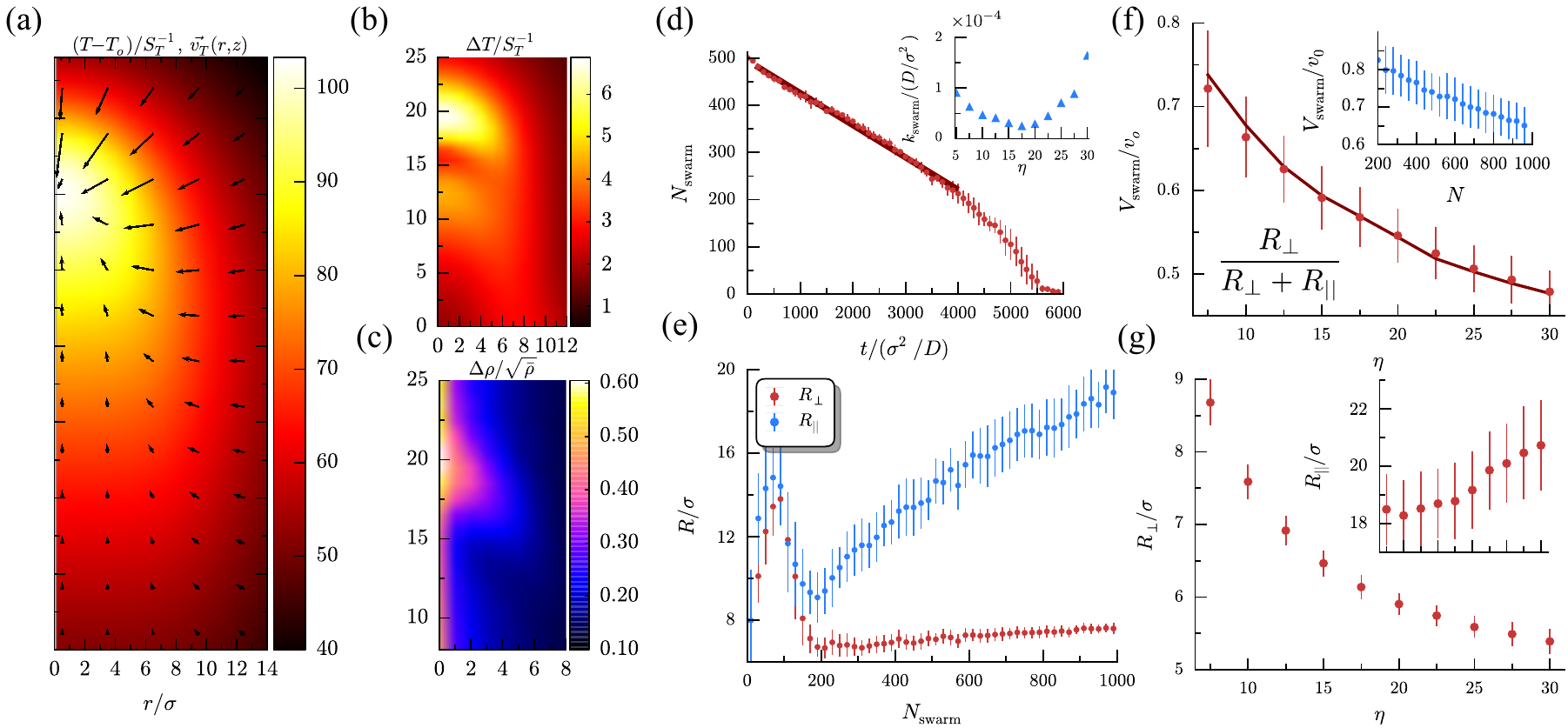}
\caption{(a) Temperature field with phoretic velocity due to temperature gradients. (b) Temperature fluctuations. (c) Density fluctuations normalized by local equilibrium expectation. (d) Time dependence of the number of colloids in the swarm for $\eta=10$ and $N_0 = 512$. This follows $N_{\rm swarm}(t) = N_0 [1 - k(\eta) t]$ until small swarm sizes with $k(\eta)$ shown inset. (e) Dependence of swarm shape, i.e. $R_\perp$ ($R_{\parallel}$) perpendicular (parallel) to illumination, on swarm size $N_{\rm swarm}$ for $\eta=10$. Dependence of swarm velocity (f) and swarm shape (g) on $\eta$ for $ 920 < N_{\rm swarm} < 980$ with solid line in (f) showing relation to the swarm shape and inset showing $N_{\rm swarm}$ dependence.  Plots showing error bars are quantities averaged over 10-20 runs with bar length showing standard deviation.\label{fig:fig2}}
\end{figure*}

We consider a collection of particles illuminated with a directed light source with uniform intensity $I$. The light intensity received at the surface of each colloid is determined by the distribution of the shadows of the colloids above it, as shown in Fig. 1a. The light received by each colloid is converted into a heat flux that increases the temperature of the colloid and the surrounding fluid in an anisotropic way. The interaction of light with particles through multiple scattering is a computationally intensive process \cite{Bohren1998}, and in order to investigate a sufficiently large number of colloids at long time scales it is desirable to introduce some simplifying features. Here we assume that the light is fully absorbed (i.e. we ignore multiple scattering) and treat the propagation of light through the colloidal dispersion via geometric optics. A particle with a clear view of the light source will have an illuminated hotter top hemisphere and a dark colder bottom hemisphere. This asymmetric temperature distribution results in the self-propulsion of the colloid via a process known as self-thermophoresis \cite{gla2007}, with a maximum velocity of magnitude $v_o=I |D_T|/(9 \kappa)$, where $D_T$ is the thermophoretic mobility (also known as the thermodiffusion coefficient) and $\kappa$ is the thermal conductivity, which is set to be equal for the colloid and the solvent for simplicity. When $D_T$ is negative, which is allowed as it is an off-diagonal Onsager coefficient \cite{de Groot} and possible via appropriate surface treatment of colloids \cite{Piazza2008}, the self-propulsion will be predominantly towards the light source with a velocity $\vec{v}_s$, leading to an effective attractive artificial phototaxis. Moreover, all colloids (whether illuminated or not) will experience a thermodiffusion drift velocity due the temperature gradient generated by the illuminated colloids, ${\vec v}_T = -D_T {\vec \nabla} T$ (see Fig. 1a). Our choice of negative $D_T$ means that the colloids act both as heat sources and heat seekers; a combination that could lead to self-organization and instability, as seen in a diverse range of nonequilibrium phenomena. These include the example of thermal explosion as first studied by D.A. Frank-Kamenetskii in 1939 \cite{DAFK1939} (and its colloidal analogue \cite{rg2012}), collective thermoregulation in a bee swarm that is used to ``cook'' their enemies \cite{MahaBees}, and the related phenomenon of pattern formation and collapse in bacterial chemotaxis \cite{KS1970,Brenner95} (and its colloidal analogue \cite{Theurkauff2012,NYU2013}).

We model the colloid as a solid core with diameter $\sigma$ and thermal conductivity $\kappa_i$ coated with a light-absorbing shell of thickness $\delta$, where we assume $\delta \ll \sigma$ (valid for nano-coatings of a microsphere) suspended in a liquid medium of thermal conductivity $\kappa_o$. We solve the steady state temperature fields on the surface of the colloid using a standard spherical harmonics expansion where an additional flux term is provided by the intensity of light, $I(\theta,\phi)$, at the colloid surface. The surface temperature  $T(\theta, \phi)$, is given by expansion coefficients $T_{l,m} = \frac{1}{2} {\sigma I_{l,m}}{/[l\kappa_i + (l+1)\kappa_o]}$, where $I_{l,m}$ is the corresponding expansion coefficient of the surface light intensity. The phoretic self-propulsion velocity of the colloid is obtained through a surface average of the slip velocity $\vec{v}_s = \frac{1}{4\pi} D_T \int \sin \theta d \theta d\phi \;\vec{\nabla}_{\parallel} T(\theta, \phi)$ \cite{gla2007}.

We study the collective behavior of the active colloids using Brownian dynamics simulation. Colloid positions are integrated through an overdamped Langevin equation scheme where positions from time step $n$ to $n+1$ are updated by ${\vec{r}}^{(n+1)} = {\vec{r}}^{(n)} + d{\vec{r}}$,  where $d{\vec{r}} = {\vec{v}}\Delta t + {\vec{\xi}}\sqrt{2\Delta t}$ using units $\sigma$ and $\sigma^2/D$ for distance and time, respectively. Random thermal fluctuation are included through the white noise term $\vec{\xi}$ with $\langle \xi_a(t) \rangle = 0$ and $\langle \xi_a(t)\xi_b(t^\prime) \rangle = \delta_{ab} \delta(t-t^\prime)$ ($a,b=1,2,3$). We neglect the change in viscosity due to heating of the water, which is known to lead to enhanced diffusion \cite{Rings2010}. The particle velocity consists of three parts $\vec{v} = \vec{v}_s + \vec{v}_T + \vec{v}_{\rm ex}$ where $\vec{v}_s$, $\vec{v}_T$ and $\vec{v}_{\rm ex}$ are the self-propulsion, collective thermal drift, and excluded volume components. The colloid self-propulsion velocity, as calculated from the average of the surface slip velocity, becomes $ \vec{v}_s = \eta \left[ \sqrt{2} \left(-{\rm Re}\left\{ {I}_{1,1}\right\} \hat{x} + {\rm Im}\left\{ {I}_{1,1}\right\} \hat{y} \right) + {I}_{1,0} \hat{z} \right]/(3\sqrt{3\pi}) $ where we set $\kappa_i=\kappa_o$ for simplicity. The collective drift velocity is given to the lowest order by $\vec{v}_{T,i} =  \frac{\eta}{8\sqrt{\pi}} \sum_{j=1}^N \frac{{I}^{(j)}_{0,0}}{|{\vec{r}}_j - {\vec{r}}_i |^2} {\hat{r}}_{ij}$ where $i$ and $j$ are particle indices and $\hat{\vec{r}}_{ij} = ({\vec{r}}_j - {\vec{r}}_i)/|{\vec{r}}_j - {\vec{r}}_i|$ is the unit vector along the center line of the two particles. This is consistent up to $\mathcal{O}(r^3)$, at which point multiparticle hydrodynamics should be included. The exclude volume component takes the Lennard-Jones form $\vec{v}_{\rm ex} = 24(2r^{-13} - r^{-7}){\hat{r}}_{ij}$ for $r<2^{1/6}$ and zero otherwise. Excluded volume and surface light intensity are calculated through the use of neighbor lists and integration is performed with an adaptive time step that constrains the maximum displacement to be less than $\sigma/100$ on average over the last $100$ steps. The behavior of the system depends on the intensity of the light source, which we can represent using the dimensionless coupling strength $\eta = \sigma I |S_T|/\kappa$, where $S_T = D_T/D$ is the Soret coefficient, and $D = k_B T/(3\pi\mu\sigma)$ is the colloid diffusion coefficient, with $\mu$ being the viscosity of the solvent. All results presented here are for simulations of up to $N_0=1024$ colloids with $\eta$ varied between $5$ and $30$. We define the swarm as the most populated set of particles with a maximum separation between particles of ${\cal L} = 10\sigma$. This distance was set by a crossover observed in the radial distribution function. We use the ordered eigenvalues $\lambda_1^2<\lambda_2^2<\lambda_3^2$ of the second central moment tensor, $S_{ab} = \sum_i^{N_{\rm swarm}} r^{(i)}_a r^{(i)}_b$, to quantify the shape of the swarm perpendicular, $R^2_{\perp} = \lambda^2_1 + \lambda^2_2$, and parallel, $R^2_{\parallel} = \lambda^2_3$, to the axis of illumination.

The colloids are initially arranged randomly in a cube of side length $20\sigma$ with the directed light source oriented along $-\hat{z}$. After a short transient period the system self-organizes into a moving swarm of $N_{\rm swarm}$ colloids with a comet like structure: a high density head region with the outer most illuminated colloids generating a central hot core, and a relatively more dilute trailing aggregate in the form of a tail; see Fig. 1a and Movie S1 in the Supplemental Material \cite{SM}.

To probe the properties of the swarm, we construct the time- and ensemble-averaged density and temperature fields in the center-of-mass moving frame of reference. The axially averaged density field is presented in Fig. 1b and the temperature field and field fluctuations can be seen in Figs. 2a and 2b, respectively. The high density head region forms a hot core which pulls the tail of the comet along and also drives the fluctuations. Normalizing the local density fluctuations by the equilibrium expectation value $ \Delta \rho/\sqrt{\rho}$ reveals spatially dependent nonequilibrium density fluctuations as shown in Fig. 2c. A particularly interesting mode of density fluctuations occurs at the tip of the head region as a result of the illuminated self-propelled particles (with the strongest $v_s$ component) attempting to escape the influence of the thermal attraction (also at its strongest). These particles usually return to the swarm, although spectacular ejection events are also observed at the tip with likelihood increasing with $\eta$; see Movie S2 for an example \cite{SM}. Density fluctuations at the swarm tip and temperature fluctuations are intertwined due to the transient appearance of heat sources.

Colloids move faster outside the swarm producing a novel circulation in the average colloid velocity streamlines in the swarm center-of-mass frame, as shown in Fig. 1b; the colloids that are attracted to the hot core can reverse their direction on crossing the shadow boundary. This phenomenon results from the competition between the strong thermally induced drift velocity $\vec{v}_T$. towards the hot core and the propulsion of individual colloids towards the light source, $\vec{v}_s$. The (potentially partially) illuminated colloids that are near the boundary of the swarm introduce a ``thermal drag'' that slows down the swarm as compared to the external fully illuminated isolated colloids.

The swarm is a long-lived but transient structure; it is subject to a slow leakage that eventually dissolves it. The number of colloids in the swarm, $N_{\rm swarm}(t)$, decreases linearly in time as $N_{\rm swarm}(t) = N_0 [1 - k(\eta) t]$ up until small ($\sim 200$ colloids) swarm sizes when dissolution occurs; see Fig. 2d. The rate of loss $k(\eta)$ is dependent on the coupling strength with a minimum occurring at $\eta \sim 17.5$; see Fig. 2d (inset). Colloids diffuse out of the shadowed tail area with some returning to the swarm and others propelling past to be permanently lost. At $\eta > 17.5$ colloids can also escape at the swarm tip due to large fluctuations in the swarm shape driven by deviations from cylindrical symmetry brought forth from colloids in the tail escaping the shadow to become active. Above the dissolution size, the swarm adopts a well-defined $R_{\perp}$, while it elongates to accommodate the given number of colloids in the swarm, as seen in Fig. 2e. As $N_{\rm swarm}$ is decreasing in time, quantities such as shape (and the swarm velocity; see below) should be measured at a fixed number.

Figure 2f shows how the thermal drag becomes more prominent as the coupling strength is increased leading to an effectively sub-linear increase of $V_{\rm swarm}$ with respect to $\eta$. Smaller swarms are seen to move faster; see Fig. 2f inset. The average shape of the swarm is also affected by the value of $\eta$, as shown in Fig. 2g. The radius perpendicular (parallel) to the axis of illumination becomes smaller (larger) as $\eta$ is increased, resulting in an increased aspect ratio. The swarm velocity is dependent on the swarm geometry as it sets the area available to receive light and the curvature. Figure 2a suggests that we can regard the boundary of the swarm as a constant-temperature surface. In analogy with perfect conductors, we can relate the number of particles receiving light in the head (tail) $N_{\rm head}$ ($N_{\rm tail}$) to the local radius of curvature, namely, $\frac{N_{\rm head}}{R_{\perp}} \sim \frac{N_{\rm tail}}{R_{\parallel}}$. We expect $V_{\rm swarm}$ to be $v_o$ less some thermal drag contribution $v_d N_{\rm tail}/N_{\rm head} \sim v_d R_{\parallel}/R_{\perp}$. Setting $v_d \sim V_{\rm swarm}$ we find $V_{\rm swarm} \sim v_o R_{\perp}/(R_\perp+R_{\parallel})$, which agrees well with the data shown in Fig. 2f.

Although we have performed our analysis using geometric optics and assuming perfection absorption, we expect many of the features of our results to remain valid under more general conditions. Any sample of light absorbing colloids illuminated with a directed light source will experience a temperature gradient along the illumination axis providing the necessary conditions for self-organization and collective propulsion. The initial configuration of the colloids before illumination needs to be sufficiently dense to initiate clustering, but otherwise is unimportant. Dimensional values for this system can be estimated by considering coated polystyrene colloids with thermal conductivity $\kappa \sim 0.1 \,{\rm W/(m\cdot K)}$, diameter $\sigma = 1\,{\rm \mu m}$, and Soret coefficient $|S_T| = 10\, {\rm K^{-1}}$, which will yield the required irradiation intensity as $I \sim 10^4 \, \eta \,{\rm W/m^2}$ in terms of $\eta$ (that gives the required laser power as $P \sim 10 \,\eta \,{\rm mW}$ for an irradiated area of $A = 1 \,{\rm mm^2}$). Using the value for the Soret coefficient, we can estimate that temperature variations will be smaller than $20\,{\rm K}$ for the highest $\eta$ presented here. The diffusion coefficient for the $\sigma = 1\,{\rm \mu m}$ colloid in water around room temperature, $D \sim 0.4 \,{\rm \mu m^2 s^{-1}}$, sets the maximum colloid velocity to $v_o \sim 0.05 \,\eta\,{\rm \mu m}/{\rm s}$, which yields typical velocities $\sim 1 \,{\rm \mu m}/{\rm s}$ for the values of $\eta$ studied here. The corresponding life time of the swarm will be $\sim 5000 \times \sigma^2/D \sim 10^4 \;{\rm s}$, which is reasonably long. The estimates suggests our results are well within reach of experiments for appropriately synthesized colloidal particles with negative Soret coefficient.

In conclusion, we have found that light-induced self-thermophoretic active colloidal particles can spontaneously self-organize to form a long-lived swarm shaped like a comet, if the colloids have a negative Soret coefficient. It is important to note that while there have been many previous studies of the collective dynamics of self-propelled particles, few (if any) have considered the details of the mechanism of propulsion, the interaction between the particles, and their collective behavior starting from underlying physical principles (and not using phenomenological models) as we have done here. Our study suggests that it is possible to design collective behavior through emergence in active colloidal suspensions. This could have applications in how smart functional materials are designed.

We would like to acknowledge fruitful discussions with M. S. Turner. This work is supported by EPSRC (JAC) and Human Frontier Science Program (HFSP) grant RGP0061/2013 (RG).

\end{document}